\documentclass[sn-mathphys]{sn-jnl}

\jyear{2021}%

\theoremstyle{thmstyleone}%
\theoremstyle{thmstyletwo}%

\theoremstyle{thmstylethree}%

\usepackage{bm}

\begin{document}

\title[Generalized Greenberger-Horne-Zeilinger arguments]{Generalized Greenberger-Horne-Zeilinger arguments from quantum logical analysis}


\author*[1]{\fnm{Karl} \sur{Svozil}}\email{svozil@tuwien.ac.at}

\affil*[1]{\orgdiv{Institute for Theoretical Physics}, \orgname{TU Wien}, \orgaddress{\street{Wiedner Hauptstra{\ss}e 8-10/136}, \city{Vienna}, \postcode{1040}, \country{Austria}}}


\abstract{The Greenberger-Horne-Zeilinger (GHZ) argument against noncontextual local hidden variables is recast in quantum logical terms of fundamental propositions, states and probabilities. Unlike Kochen-Specker- and Hardy-like configurations, this operator based argument proceeds within four nonintertwining contexts. The nonclassical performance of the GHZ argument is due to the choice or filtering of observables with respect to a particular state. We study the varieties of GHZ games one could play in these four contexts, depending on the chosen state of the GHZ basis.}

\keywords{Greenberger-Horne-Zeilinger argument, Gleason theorem, Kochen-Specker theorem, Born rule, gadget graphs, Greechie diagram, McKay-Megill-Pavicic diagram (MMP), orthogonality hypergraph}



\maketitle

\setcounter{MaxMatrixCols}{20}

\section{Quantum logical structures}

In what follows, the Greenberger-Horne-Zeilinger (GHZ) argument~\cite{ghz}
will be recast in purely quantum logical terms.
In particular, the operators corresponding to GHZ observables will be written in their spectral form,
such that the respective orthogonal projection operators can be identified with elementary propositions.
Since the argument is state-dependent, the appropriate states need to be identified,
and their predictions and expectations on those elementary propositions need to be evaluated.
This facilitates the discernment of important structural components.
Such insights can then be used to embed and relate GHZ to traditional quantum logical findings,
as well as to generalize and extend the argument to more general experimentally verifiable predictions and assertions.

\subsection{Appraisal}

The configuration of observables and states that constitute a
GHZ argument
renders a capacity to perform on specific tasks (although it is worse for other tasks, see Section~\ref{2020-ghz-gtcbwbcbnbqm})
that no classical means can achieve:
It lets three parties ``win'' a particular task with certainty,
whereas there is only a probabilistic classical chance to do so classically.

The situation is not unlike the quantum Deutsch algorithm~\cite[Section~2.2]{mermin-07}:
It yields a particular property (parity) of the input with one query with certainty,
whereas there is only a probabilistic classical chance to do so classically.

There are potential drawbacks though since this performance increase is not ``universal''.
Because on the one hand, other tasks can only be achieved by invoking different states and operators;
and, on the other hand, ``complete'' determination of some unknown
quantum state (without preselection) requires the same number of queries as in the classical case.
As a result, the capacity of such algorithms is problem and query-specific.
Yet, as long as one is interested in increasing capabilities or performance
not universally but for a particular task or ``game''
such quantum advantages exist and are exploitable.

What ``drives'' GHZ is quantum entanglement: the capacity to relationally encode~\cite{schrodinger-gwsidqm2,zeil-99} multi- (in this case three-) partite systems
in such a way that the (information of the) properties of the constituents are defined merely as their collective behavior.
Again, there is a price involved, in this case, the complete loss of (information of the) properties of the individual constituents.
Indeed, for pure states individual and relational properties can be rewritten into each other by nonlocal ``scrambling''
unitary transformations ``sampling'' individual into relational states~\cite{svozil-2016-sampling}.
Such entangled states can ``carry''  relational information about collective properties of their constituents that no classical state can.

\subsection{Motivation for logico-algebraic analysis}

As mentioned earlier the GHZ argument will be very explicitly analyzed in terms of its quantum logical aspects;
in particular, concerning the observables and states involved.
One advantage of such an analysis is accessibility to other fields of research,
as it appears less ad hoc and complementing existing research.

It might not be too unreasonable to state that, for classical-versus-quantum discord, the GHZ argument ``competes'' with
the  Kochen-Specker (KS) theorem~\cite{kochen1,specker-ges},
even though the latter can be converted into the former~\cite{Xu-Chen-Guehnw-2020,cabello2020converting}.
Whereas the KS theorem, as well as its weaker probabilistic form,
the Hardy-type argument~\cite{Hardy-93,svozil-2020-hardy} (sometimes referred to as 1-implies-$\{0,1\}$~\cite{svozil-2006-omni} or true-implies-\{false.true\}-rule~\cite{2018-minimalYIYS})
often is ``local'', GHZ uses a particular three-partite configuration whose constituents can be space-like separated.

KS and Hardy-type configurations use elementary dichotomic yes-no or 0--1 propositions
(which can be encoded as a normalized vector $\vert {\bf a} \rangle$ (in three or higher dimensions)
spanning a one-dimensional subspace of the Hilbert space,
and which is equivalent to the orthogonal projection operator $\vert {\bf a} \rangle \langle {\bf a}\vert $
projecting the Hilbert space onto this subspace).
In contradistinction, GHZ use expectations of three dichotomic variables, each of which assumes either the value 1 or -1,
and triple distributions.

Thereby, the GHZ argument is operator-based~\cite{Holweck2017},
because it allows observables and their associated self-adjoint operators with eigenvalues
``$+1$'' and ``$-1$''---that should be understood in terms of their spectral forms as functions of orthogonal projection operators
occurring in the KS and Hardy-type configurations.
(Functions of normal operators can be defined  by their spectral form~\cite[\S~82, p.~165-169]{halmos-vs}.)
In what follows we shall also find that, whereas the KS argument involves many intricately intertwining contexts,
the~GHZ argument uses a single isolated context.

\section{Classical GHZ games}

Routine approaches to classical-versus-quantum discord start from some collection of quantum observables
and attempt to force classical interpretations upon them.
In contrast to this manners the GHZ argument will be motivated by
the lack of capacity to perform certain tasks by classical means,
whereas quantum capacities to achieve these tasks exist.
The presentation starts with an exposition by
Bacon (aka ``The Quantum Pontiff'')~\cite{bacon-ghzgames-2006} (see also Refs.~\cite{Broadbent-PRNLboxes,Scarani-prboxes})
in which certain quantum resources allow players always to win,
whereas this cannot be guaranteed classically.

Suppose some prison ward allows three parties to pre-select some ``share'', and subsequently isolates them in separate cells
(without further communication).
Afterward, the ward distributes paper slips to each one of them, such that each slip contains a single symbol out of two symbols -- say, ``$x$'' or ``$y$'' -- per party and slip,
in either one of the following four configurations or contexts: $xxx$, $xyy$, $yxy$, and $yyx$ (the order reflects the order of the parties).
In particular, the ward does not reveal which type of configuration is chosen,
so from the intrinsic ``local'' perception of the single isolated prisoners,
upon reception of ``$x$'' or ``$y$'' it could be two different configurations or contexts
(the possibility of the other two configurations can be eliminated by exclusion).
Upon receiving the slip, all three parties must then write on their slips (or shout) simultaneously and without any further coordination either ``$-$''1 or ``$+$''1.

\subsection{Positive products from squares of outcomes}
Suppose the goal is, in algebraic terms, to form negative products of these three factors for all four configurations $xxx$, $xyy$, $yxy$, and $yyx$.
Does there exist a ``classical strategy'' for those players to always win? Yes, because by a parity argument, if one forms
$
\underbrace{(x \cdot x\cdot x)}_{-1}
\cdot
\underbrace{(x\cdot y\cdot y)}_{-1}
\cdot
\underbrace{(y\cdot x\cdot y)}_{-1}
\cdot
\underbrace{(y\cdot y\cdot x)}_{-1} = (-1)^4=1$---and
thereby multiplies all factors of all of the four configurations---one
obtains $x^6 y^6$ which allows a classical strategy of writing (or shouting)
``$-$''1 on $x$, and anything (but previously coordinated and fixed), that is, ``$\pm$''1 on $y$.
More explicitly, suppose the prisoners have agreed to write (or shout) ``$-1$'' for $x$ and ``$+1$'' for $y$;
then
$
\underbrace{(\overbrace{x}^{-1} \cdot \overbrace{x}^{-1}\cdot \overbrace{x}^{-1})}_{-1}
=
\underbrace{(\overbrace{x}^{-1}\cdot \overbrace{y}^{+1}\cdot \overbrace{y}^{+1})}_{-1}
=
\underbrace{(\overbrace{y}^{+1}\cdot \overbrace{x}^{-1}\cdot \overbrace{y}^{+1})}_{-1}
=
\underbrace{(\overbrace{y}^{+1}\cdot \overbrace{y}^{+1}\cdot \overbrace{x}^{-1})}_{-1}$
and the prisoners always win.
Very similar considerations apply to requests of the prison ward to produce only positive results,
or more general demands that will be discussed later.

\subsection{GHZ goal: negative products from squares of outcomes}
However, some goals for the prisoners are unachievable by classical means.
In particular, let us specify the GHZ game as the goal set by the prison ward as follows:
``produce the products
$x \cdot x\cdot x = +1$ and  $x\cdot y\cdot y=y\cdot x\cdot y=y\cdot y\cdot x=-1$.''
Then, by the same parity argument mentioned earlier, the prisoners cannot win all the time, since then the combined product
$
\underbrace{(x \cdot x\cdot x)}_{+1}
\cdot
\underbrace{(x\cdot y\cdot y)}_{-1}
\cdot
\underbrace{(y\cdot x\cdot y)}_{-1}
\cdot
\underbrace{(y\cdot y\cdot x)}_{-1}
=(+1)(-1)^3=-1$
would be negative but the respective factors occur in multiples of squares $x^2$ and $y^2$ and
must therefore result in a positive product---a perfect contradiction.
Note that the earlier strategy of writing (or shouting) ``$-$''1 on $x$, and anything, that is, ``$\pm$''1 on $y$
would still succeed whenever the ward invokes slips that carry $xyy$, $yxy$, and $yyx$ on them.
But this strategy fails miserably in the positive $xxx$ case.
Conversely, the ``inverse'' strategy  of writing (or shouting) ``$+$''1 on $x$, and anything, that is, ``$\pm$''1 on $y$
would succeed in the positive $xxx$ case
but fail for all other cases requesting negative products $xyy$, $yxy$, and $yyx$.

Note that if the ward would also reveal the configuration chosen then all of these games could easily be won by the prisoners
with classical means,
albeit not in a context-invariant or context-independent way.
Because then, from each of the four classical configurations or contexts, they could choose a single instance that would fit
the game---say, $x_+x_+x_+$ for the $xxx$ configuration, and $x_-y_+y_+$, $y_+x_-y_+$, and $y_+y_+x_-$ for the other three configurations---and thereby
win the earlier game not recoverable by classical noncontextual means.
Here the context dependence is in the $x$ assignments: ``$+$''1 for the $xxx$ configuration, ``$-$''1
for the other configurations $xyy$, $yxy$, and $yyx$.

However, this nondisclosure of configurations and contexts is not how the GHZ game has been operationalized and empirically presented~\cite{PhysRevLett.82.1345,panbdwz}.
There, the final configuration or context $xxx$ [which supposedly serves as a criterion for (non)classicality]
is not subjected to a ``delayed-choice'' involving space-like separated events.
Therefore, this performance could in principle also be rendered classically by noncontextual value assignments discussed later.

Nevertheless, the original GHZ protocol offers a quantum method for the prisoners to win all the time---that is, for all cases
the ward may choose---without disclosing the chosen configuration or context to the prisoners.
But before going into this particular realization, the classical case will be formalized a little further.

\subsection{Contexts involved in classical GHZ configurations}

We shall argue that the classical Shimony-Mermin form~\cite{ghsz,mermin90b,mermin} of the GHZ argument employs four separate nonintertwining ``complementary'' contexts.
For the sake of the argument suppose one is dealing with classical objects, such as generalized urn models~\cite{wright} or finite automata~\cite{svozil-2001-eua,svozil-2018-b}
which encode (a classical form of) complementarity.
Alternatively one might think of just classical objects which can have two dichotomic properties, namely
the dichotomic observables
\begin{equation}
{{x}} \in \{x_+,x_-\}     \textrm{, and }  {{y}} \in \{y_+,y_-\}
,
\label{2020-ghz-hl}
\end{equation}
where $x_+$ and $x_-$ as well as $y_+$ and $y_-$ are the states corresponding to outcomes of measurements of ${{x}}$ and ${{y}}$, respectively.

Suppose further that one is drawing triples of identical balls from a generalized urn.
Consider four types of measurements on the first, second and third ball:
\begin{equation}
 \{  {{x}} , {{x}} , {{x}}  \}
,
 \{  {{x}} , {{y}} , {{y}}  \}
,
 \{  {{y}} , {{x}} , {{y}}  \}
, \textrm{ and }
 \{  {{y}} , {{y}} , {{x}}  \}
.
\label{2020-ghz-ghlm}
\end{equation}

\begin{figure}[htb]
\begin{center}
\begin{tabular}{ c }
\includegraphics{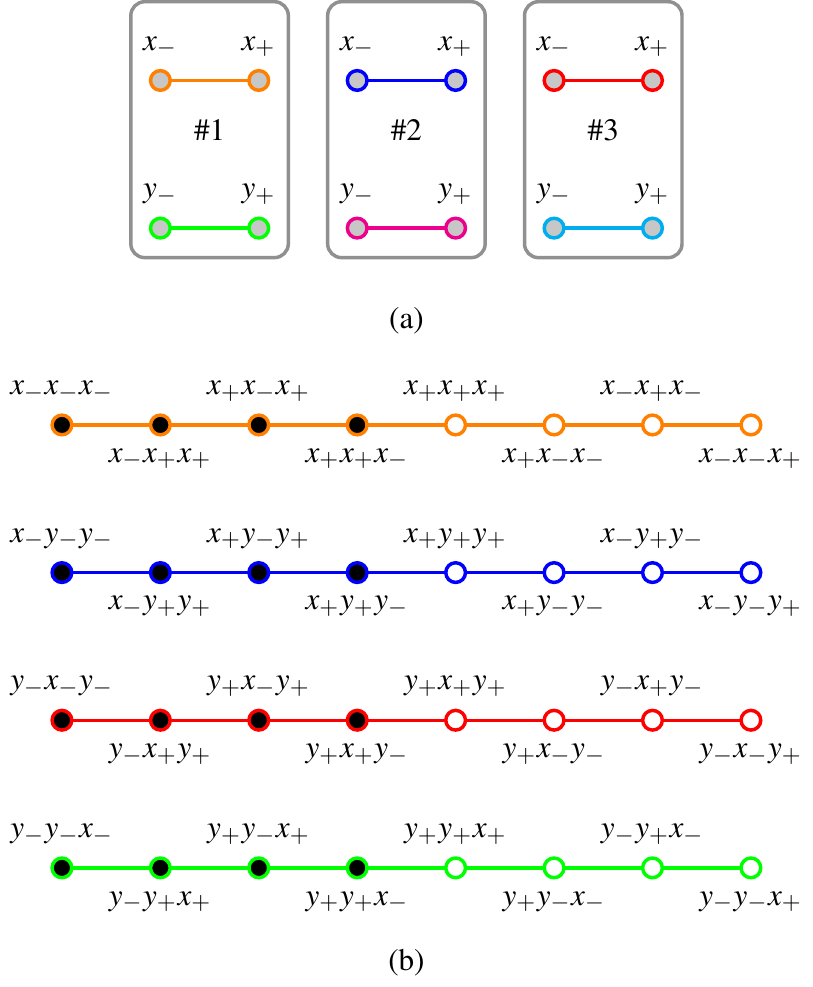}
\end{tabular}
\end{center}
\caption{\label{2020-f-ghz-contextconf}
Hypergraphs representing
(a) the six disconnected classical single-particle contexts representing the observables of the first, second, and third particle, respectively;
(b) four isolated, nonintertwining contexts, with eight atoms each, employed in the GHZ argument
(there are eight contexts in total, but only half of them are considered).
Filled circles indicate states which are classically allowed in a modified GHZ game
requesting negativity for all products  $xyy=yxy=yyx=xxx=-1$.
Every column represents a viable noncontextual winning strategy.
}
\end{figure}

In what follows, we shall use (conformal, orthogonality) hypergraphs~\cite[Sect.~2.4]{Bretto-MR3077516} introduced by Greechie~\cite[p.~120]{greechie:71}
that depict
logical configurations or
contexts as smooth lines~\cite{Greechie1968,kalmbach-83,svozil-tkadlec,Mckay2000,Pavicic-2005,Bretto-MR3077516,2018-minimalYIYS}.
The term hypergraph should be understood in the broadest possible consistent sense.
Figure~\ref{2020-f-ghz-contextconf}(a) represents the six Boolean algebras
associated with the propositional structures of six single-particle observables:
there are three particles, each measured at two angles or directions, respectively.
Simultaneous ``triple'' measurements of one (of two, either $x$ or $y$) observable per particle, on the three particles,
result in $2^3=8$ Cartesian product triples
forming the configurations or contexts
$\{x,x,x\}$,
$\{x,x,y\}$,
$\ldots$,
$\{y,y,y\}$.
Only four of these configurations or contexts, namely
$\{x,x,x\}$,
$\{x,y,y\}$,
$\{y,x,y\}$, and
$\{y,y,x\}$,
are employed in the GHZ argument.
Figure~\ref{2020-f-ghz-contextconf}(b) represents the eight possibilities, cases or instantiations per type of measurement
of the latter four GHZ contexts
(enumerated in lexicographic order; for brevity
we shall write ``${{a}}{{b}}{{c}}$''
for the ordered triple ``$\left[ {{a}},{{b}},{{c}} \right]$'').
\begin{equation}
\begin{aligned}
C_\textrm{c,GHZ}^{{{x}}{{x}}{{x}}}=& \{ x_+x_+x_+,x_+x_+x_-,x_+x_-x_+,x_+x_-x_-,\\ &\qquad x_-x_+x_+,x_-x_+x_-,x_-x_-x_+,x_-x_-x_- \}
,    \\
C_\textrm{c,GHZ}^{{{x}}{{y}}{{y}}}=& \{ x_+y_+y_+,x_+y_+y_-,x_+y_-y_+,x_+y_-y_-,\\ &\qquad x_-y_+y_+,x_-y_+y_-,x_-y_-y_+,x_-y_-y_- \}
,        \\
C_\textrm{c,GHZ}^{{{y}}{{x}}{{y}}}=& \{ y_+x_+y_+,y_+x_+y_-,y_+x_-y_+,y_+x_-y_-,\\ &\qquad y_-x_+y_+,y_-x_+y_-,y_-x_-y_+,y_-x_-y_- \}
, \textrm{ and}  \\
C_\textrm{c,GHZ}^{{{y}}{{y}}{{x}}}=& \{ y_+y_+x_+,y_+y_+x_-,y_+y_-x_+,y_+y_-x_-,\\ &\qquad y_-y_+x_+,y_-y_+x_-,y_-y_-x_+,y_-y_-x_- \}
.
\end{aligned}
\label{2020-ghz-ghlmcontexpl}
\end{equation}
which, bundeled into a set, are identified with the set of four~GHZ contexts
\begin{equation}
 \Big\{
C_\textrm{c,GHZ}^{{{x}}{{x}}{{x}}}
,
C_\textrm{c,GHZ}^{{{x}}{{y}}{{y}}}
,
C_\textrm{c,GHZ}^{{{y}}{{x}}{{y}}}
,
C_\textrm{c,GHZ}^{{{y}}{{y}}{{x}}}
\Big\}
.
\label{2020-ghz-clghzc}
\end{equation}

Suppose that $x_+$, $x_-$, $y_+$ and $y_-$ are real nonvanishing numbers.
The outcomes are multiplied, such that $[a,b,c]=a\times b\times c$.
Multiplication of the observables in the four contexts results in a bound from below:
\begin{equation}
({{x}}{{x}}{{x}})
({{x}}{{y}}{{y}})
({{y}}{{x}}{{y}})
({{y}}{{y}}{{x}})
 =
({{x}}{{y}})^6 >0.
\label{2020-ghz-ghlmcont}
\end{equation}
In particular, if one identifies $x_+=y_+=1$ and $x_-=y_-=-1$ then $[({{x}}{{y}})^3]^2 =1$.
As already pointed out earlier, this classical prediction rests upon a parity argument:
any square, indeed any even exponent, of a nonzero real number is positive.
Note also that, by taking the product of the first three terms,
one obtains $\left[(\pm x)^2(\pm y)^2\right]^3={{y}}^6  {{x}}^3 =  {{x}}^3 $.
This should always be identical to the last term  in~(\ref{2020-ghz-ghlmcont}),
which is ${{x}}{{x}}{{x}}$.
Because according to the classical assumption of noncontextuality---independence of the observed value of ${{x}}$
on whatever other observables are measured simultaneously alongside it in different configurations or contexts---and
value definiteness, it should not make any difference
in which of the four contexts in~(\ref{2020-ghz-ghlmcont})
${{x}}$ is measured.
This, and the equivalent bound from below
in~(\ref{2020-ghz-ghlmcont}), are the classical predictions.

The classical probabilities of this configuration are the convex summations of $4^3$ extreme cases
from the 4 individual cases
$\{x_+y_+,x_+y_-,x_-y_+,x_-y_-\}$
per one of the three particles.
If interpreted geometrically  these $4^3$ vertices in six-dimensional space,
and their equivalent representations in terms of inequalities of the hull of the
convex polytope spanned by them, has been enumerated for three-partite correlations~\cite{2000-poly}.
In all $4^3$ classical cases, and for all classical probability distributions constructed from them, prediction~(\ref{2020-ghz-ghlmcont}) is satisfied.

Note that the classical GHZ contexts are isolated from each other and do not ``communicate''
via common intertwine observables.
As has been pointed out earlier,
in this aspect, they differ from other KS or Hardy-type arguments involving gadgets of intertwining contexts.

\section{Quantum advantage}

Let us now turn to quantum mechanical predictions of quantum doubles of the classical configuration of contexts introduced earlier;
in particular to a GHZ configuration that yields a winning strategy to the classical GHZ game
goal set by the prison ward as follows:
``produce the products
$x \cdot x\cdot x = +1$ and  $x\cdot y\cdot y=y\cdot x\cdot y=y\cdot y\cdot x=-1$.''
This is a rewrite of a confifuration that was presented
by Greenberger, Horne and Zeilinger~\cite{ghz},
later with Shimony~\cite{ghsz}.
(I still vividly remember that earlier Zeilinger
had suggested to me to pursue  ``a three-particle analog of the Bell-type two-particle setup'' but I had cordially abstained.)
Immediately afterward Mermin gave a uniform presentation of the GHZ argument~\cite{mermin90b,mermin1}
that can readily be rewritten in terms of the underlying elementary propositions,
a task which is pursued here.
Subsequently, these quantum predictions have been empirically confirmed~\cite{PhysRevLett.82.1345,panbdwz};
albeit with some provisos which will be discussed later.

By a quantum realization is some identification of (i) the observables ${{x}}$ and ${{y}}$
with self-adjoint operators; in particular, with two-times-two Hermitian matrices with a dichotomic spectrum $\{-1,+1\}$,
and (ii) a pure state identified with a nonzero (unit) vector, or its associated one-dimensional orthogonal projector,
or the one-dimensional subspace spanned by the vector.
Only pure states which are associated with normalized vectors are considered; that is, we shall not consider mixed states.

For historic and empirical reasons---for instance, the experimental realizations in terms of spin-$\frac{1}{2}$
particles--particular quantum realizations are often written in terms of Pauli spin matrices.
The general form of the Pauli spin matrices in spherical coordinates is given by
$
\bm{\sigma}( \theta ,\varphi )= \sigma_x \sin\theta \cos\varphi  + \sigma_y \sin\theta \sin\varphi  + \sigma_z \cos\theta =
\begin{pmatrix} \cos \theta  &e^{-i\varphi} \sin \theta   \\
  e^{i\varphi}\sin \theta  & -\cos \theta
  \end{pmatrix}
$, where $0 \le \theta \le \pi$ is the polar angle in the $x$--$z$-plane
from the $z$-axis,
and $0 \le \varphi < 2 \pi$ is the azimuthal angle in the $x$--$y$-plane
from the $x$-axis.
The usual form of the Pauli spin matrices is recovered by identifying
$
\sigma_x = \bm{\sigma} \left(\frac{\pi}{2},0\right)= \mathrm{antidiag}\begin{pmatrix} 1 , 1  \end{pmatrix}
$,
$
\sigma_y = \bm{\sigma} \left(\frac{\pi}{2},\frac{\pi}{2}\right)= \mathrm{antidiag}\begin{pmatrix} -i ,   i  \end{pmatrix}
$,
and
$
\sigma_z = \bm{\sigma} \left(0,0\right)= \mathrm{diag} \begin{pmatrix} 1 ,  -1   \end{pmatrix}
$.

\section{Quantum realization}

Mermin's Ansatz~\cite{mermin} is the following quantum realization~(\ref{2020-ghz-hl}):
\begin{equation}
\begin{aligned}
&{{y}}  \equiv \sigma_y  ,
{{x}}  \equiv \sigma_x  , \textrm{ and a shared state} \\
&\vert \Upsilon_1  \rangle  \langle   \Upsilon_1  \vert \textrm{ for the sake of GHZ discord}
.
\end{aligned}
\label{2020-ghz-hlqm}
\end{equation}

As we shall see the quantum state of the GHZ argument operates within a single context.

\subsubsection{Operators}
The observables (formalized by eight-dimensional operators) operate within four contexts;
very similar to the classical contexts discussed earlier.
These contexts can be obtained by considering the spectral forms
of the following
four mutually commuting tensor products of operators (tensor products are often denoted by the symbol ``$\otimes$''
but, for reasons of brevity, we shall not write this symbol explicitly.
This should not be confused with the matrix or dot product ``$\cdot$'';
so that, for instance,
$\sigma_y \sigma_y \sigma_x\equiv \sigma_y \otimes \sigma_y \otimes \sigma_x$):
\begin{equation}
\begin{aligned}
&\sigma_y \sigma_y \sigma_x ,\;
\sigma_y \sigma_x \sigma_y ,\;
\sigma_x \sigma_y \sigma_y
\textrm{, and }
\sigma_x \sigma_x \sigma_x \text{, with}\\
&\quad \sigma_y \sigma_y \sigma_x = \textrm{antidiag} \begin{pmatrix} -1, -1, 1, 1, 1, 1, -1, -1 \end{pmatrix},\\
&\quad \sigma_y \sigma_x \sigma_y = \textrm{antidiag} \begin{pmatrix} -1, 1, -1, 1, 1, -1, 1, -1 \end{pmatrix},\\
&\quad \sigma_x \sigma_y \sigma_y = \textrm{antidiag} \begin{pmatrix} -1, 1, 1, -1, -1, 1, 1, -1 \end{pmatrix}
\textrm{, and } \\
&\quad \sigma_x \sigma_x \sigma_x = \textrm{antidiag} \begin{pmatrix} 1, 1, 1, 1, 1, 1, 1, 1 \end{pmatrix}
.
\end{aligned}
\label{2020-ghz-sigmas-contextop}
\end{equation}
Because the operators in~(\ref{2020-ghz-sigmas-contextop}) mutually commute
the spectral forms
of
$\sigma_y \sigma_y \sigma_x$,
$\sigma_y \sigma_x \sigma_y$,
$\sigma_x \sigma_y \sigma_y$,
and
$\sigma_x \sigma_x \sigma_x$
can be written as sharing the same eight
orthogonal projection operators~\cite[Sections~79,84]{halmos-vs}.
Note that since the four-fold multiplicity of the two eigenvalues $-1$ and $1$ per operator,
its degenerate spectrum---enumerated in the columns of Table~\ref{2020-hardy-t-eveeva}---extends over only two orthogonal projection operators into four-dimensional subspaces.
The projection operators can, for instance, be directly obtained by a Lagrange polynomial of the form
$\textsf{\textbf{E}}_{\pm}  ( \sigma_y \sigma_y \sigma_x ) = \frac{1}{2}\left(\sigma_y \sigma_y \sigma_x \pm \mathbb{I}_8\right)$,
$\textsf{\textbf{E}}_{\pm}  ( \sigma_y \sigma_x \sigma_y ) = \frac{1}{2}\left(\sigma_y \sigma_x \sigma_y \pm \mathbb{I}_8\right)$,
$\textsf{\textbf{E}}_{\pm}  ( \sigma_x \sigma_y \sigma_y ) = \frac{1}{2}\left(\sigma_x \sigma_y \sigma_y \pm \mathbb{I}_8\right)$, and
$\textsf{\textbf{E}}_{\pm}  ( \sigma_x \sigma_x \sigma_x ) = \frac{1}{2}\left(\sigma_x \sigma_x \sigma_x \pm \mathbb{I}_8\right)$.
A ``refined'' resolution by one-dimensional projection operators can be achieved
by forming products of these projection operators which serve as ``filters''~\cite{DonSvo01}.

\subsubsection{Preselected states}
These eight mutually orthogonal one-dimensional projection operators correspond to the system of mutually orthogonal eigenvectors
which can be identified with the GHZ basis~\cite{mermin,krenn1,Uchida-2015} which, in turn, is identified with
the quantum mechanical GHZ-context
\begin{equation}
C_\text{q,GHZ}=\{
\vert \Upsilon_1 \rangle ,
\vert \Upsilon_2 \rangle ,
\vert \Upsilon_3 \rangle ,
\vert \Upsilon_4 \rangle ,
\vert \Upsilon_5 \rangle ,
\vert \Upsilon_6 \rangle ,
\vert \Upsilon_7 \rangle ,
\vert \Upsilon_8 \rangle
\}.
\label{2020-ghz-context}
\end{equation}
It might be more appropriate to call this the ``GHZM'' context because Merim suggested to use $\vert \Upsilon_1 \rangle$ for a GHZ argument~\cite{mermin}.
But being aware of Stigler's law of eponymy~\cite{Stigler1980,Stigler-sott}, stating that
{\it ``no scientific discovery is named after its original discoverer''},
and having encountered similar issues already earlier with the term ``Hardy-like'',
I shall leave it at that.

The components of the vectors of the GHZ basis can be expressed in terms of the Cartesian standard basis which
coincides with the set of eigenvectors of $\sigma_z = \bm{\sigma} (0,0)$:
we denote these eigenvectors by (the symbol ``$\intercal$'' indicates transposition)
$\vert z_+ \rangle =
\begin{pmatrix}1,0\end{pmatrix}^\intercal$
and
$\vert z_- \rangle  =
\begin{pmatrix}0,1\end{pmatrix}^\intercal$
corresponding to the eigenvalues $+1$ and $-1$, respectively.
Then relative to the eigenvectors of $\sigma_z = \bm{\sigma} (0,0)$
the eight eigenvectors of the four contexts
$\sigma_y \sigma_y \sigma_x$ , $\sigma_y \sigma_x \sigma_y$, $\sigma_x \sigma_y \sigma_y$, and $\sigma_x \sigma_x \sigma_x$ can be written as:
\begin{equation}
\begin{aligned}
\vert \Upsilon_1 \rangle =& \frac{1}{\sqrt{2}}\left( \vert z_+z_+z_+ \rangle  + \vert z_-z_-z_- \rangle \right) ,\;
\vert \Upsilon_2 \rangle =  \frac{1}{\sqrt{2}}\left( \vert z_+z_+z_+ \rangle  - \vert z_-z_-z_- \rangle \right) , \\
\vert \Upsilon_3 \rangle =& \frac{1}{\sqrt{2}}\left( \vert z_+z_+z_- \rangle  + \vert z_-z_-z_+ \rangle \right) , \;
\vert \Upsilon_4 \rangle =  \frac{1}{\sqrt{2}}\left( \vert z_+z_+z_- \rangle  - \vert z_-z_-z_+ \rangle \right) , \\
\vert \Upsilon_5 \rangle =& \frac{1}{\sqrt{2}}\left( \vert z_+z_-z_+ \rangle  + \vert z_-z_+z_- \rangle \right) ,\;
\vert \Upsilon_6 \rangle =  \frac{1}{\sqrt{2}}\left( \vert z_+z_-z_+ \rangle  - \vert z_-z_+z_- \rangle \right) , \\
\vert \Upsilon_7 \rangle =& \frac{1}{\sqrt{2}}\left( \vert z_+z_-z_- \rangle  + \vert z_-z_+z_+ \rangle \right) , \;
\vert \Upsilon_8 \rangle =  \frac{1}{\sqrt{2}}\left( \vert z_+z_-z_- \rangle  - \vert z_-z_+z_+ \rangle \right) .
\end{aligned}
\label{2020-ghz-evsisjsk}
\end{equation}

\subsubsection{Selection for GHZ discord}

For the sake of the original GHZ discord, and thereby to produce
a contradiction with the classical predictions~(\ref{2020-ghz-ghlmcont}),
we shall desire---and choose,
if possible, a ``suitable'' element $\vert \Upsilon_i \rangle $ of $C_\text{q,GHZ}$
which by construction is one of the eigenstates of the operators in~(\ref{2020-ghz-sigmas-contextop})---for which
\begin{equation}
\begin{aligned}
\sigma_y \sigma_y \sigma_x \vert \Upsilon_i \rangle &=
\sigma_y \sigma_x \sigma_y \vert \Upsilon_i \rangle =
\sigma_x \sigma_y \sigma_y \vert \Upsilon_i \rangle = - \vert \Upsilon_i \rangle\textrm{, and } \\
\sigma_x \sigma_x \sigma_x \vert \Upsilon_i \rangle &= + \vert \Upsilon_i \rangle .
\end{aligned}
\label{2020-ghz-sigmas-contextopwid}
\end{equation}
Because in this case the product of the eigenvalues $(- 1)^3(+ 1)$  becomes negative
we obtain a complete discord or contradiction with any classical prediction~(\ref{2020-ghz-ghlmcont}), as desired.

\begin{table}[t]
 \begin{center}
 \caption{\label{2020-hardy-t-eveeva}  Eigenvalues $+ \equiv +1$ and $- \equiv -1$ associated with eigenvectors for the four contexts
$\sigma_y \sigma_y \sigma_x$, $\sigma_y \sigma_x \sigma_y$, $\sigma_x \sigma_y \sigma_y$,  and $\sigma_x \sigma_x \sigma_x$
in Eq.~(\ref{2020-ghz-sigmas-contextop}).
The components of the GHZ states are relative to the Cartesian standard basis which contains the eigenstates of $\sigma_z\sigma_z\sigma_z$.
I will argue later that each of the eight rows corresponds to a particular GHZ game that is in complete discord with its classical predictions.}
 \begin{tabular}{ c c c c c c c c c c }
\toprule
GHZ state &
$\sigma_y \sigma_y \sigma_x$ & $\sigma_y \sigma_x \sigma_y$ &  $\sigma_x \sigma_y \sigma_y$  & $\sigma_x \sigma_x \sigma_x$
\\
\midrule
$\vert \Upsilon_1 \rangle = \frac{1}{\sqrt{2}}\begin{pmatrix} 1, 0, 0, 0, 0, 0, 0, 1 \end{pmatrix}^\intercal $ & {\Large -} & {\Large -} &  {\Large -} & {\Large +}\\
$\vert \Upsilon_2 \rangle = \frac{1}{\sqrt{2}}\begin{pmatrix} 1, 0, 0, 0, 0, 0, 0, -1 \end{pmatrix}^\intercal$ & {\Large +} & {\Large +} &  {\Large +} & {\Large -}\\
$\vert \Upsilon_3 \rangle = \frac{1}{\sqrt{2}}\begin{pmatrix} 0, 1, 0, 0, 0, 0, 1, 0 \end{pmatrix}^\intercal $ & {\Large -} & {\Large +} &  {\Large +} & {\Large +}\\
$\vert \Upsilon_4 \rangle = \frac{1}{\sqrt{2}}\begin{pmatrix} 0, 1, 0, 0, 0, 0, -1, 0 \end{pmatrix}^\intercal$ & {\Large +} & {\Large -} &  {\Large -} & {\Large -}\\
$\vert \Upsilon_5 \rangle = \frac{1}{\sqrt{2}}\begin{pmatrix} 0, 0, 1, 0, 0, 1, 0, 0 \end{pmatrix}^\intercal $ & {\Large +} & {\Large -} &  {\Large +} & {\Large +}\\
$\vert \Upsilon_6 \rangle = \frac{1}{\sqrt{2}}\begin{pmatrix} 0, 0, 1, 0, 0, -1, 0, 0 \end{pmatrix}^\intercal$ & {\Large -} & {\Large +} &  {\Large -} & {\Large -}\\
$\vert \Upsilon_7 \rangle = \frac{1}{\sqrt{2}}\begin{pmatrix} 0, 0, 0, 1, 1, 0, 0, 0 \end{pmatrix}^\intercal $ & {\Large +} & {\Large +} &  {\Large -} & {\Large +}\\
$\vert \Upsilon_8 \rangle = \frac{1}{\sqrt{2}}\begin{pmatrix} 0, 0, 0, 1,- 1, 0, 0, 0 \end{pmatrix}^\intercal$ & {\Large -} & {\Large -} &  {\Large +} & {\Large -}\\
\botrule
\end{tabular}
 \end{center}
 \end{table}
As mentioned earlier, the eigenvalues associated with the eigenvectors
for different contexts differ from each other but are highly multiplicitous: four eigenvectors are associated with
the same eigenvalues
$+1$ and $-1$, with  multiplicities equal to four, respectively.

Table~\ref{2020-hardy-t-eveeva} enumerates these degenarcies.
By contemplating this tabulation it is not too difficult
to single out the state~\cite{mermin}
of the GHZ basis which satisfies the discord criterium~(\ref{2020-ghz-sigmas-contextopwid}):
it is  $\vert \Upsilon_1 \rangle = \frac{1}{\sqrt{2}}\left( \vert z_+z_+z_+ \rangle  + \vert z_-z_-z_- \rangle \right)=\frac{1}{\sqrt{2}}\begin{pmatrix} 1, 0, 0, 0, 0, 0, 0, 1 \end{pmatrix}^\intercal$.

\subsubsection{General preselection state}

Let us, for the sake of completeness and for later use, write
all eight eigenstates~(\ref{2020-ghz-evsisjsk})
as well as all coherent superpositions thereof into a single closed form:
\begin{equation}
\begin{aligned}
& \vert \Upsilon  \rangle = \sum_{i=1}^8 \alpha_i \vert \Upsilon_i \rangle =
\frac{1}{\sqrt{2}} \Big( \alpha_1+\alpha_2,\alpha_3+\alpha_4,\alpha_5+\alpha_6,\\
  &\qquad \quad \alpha_7+ \alpha_8,\alpha_7-\alpha_8,\alpha_5-\alpha_6,\alpha_3-\alpha_4,\alpha_1-\alpha_2 \big)^\intercal ,
\end{aligned}
\label{2020-ghz-psiGHZ}
\end{equation}
with $\sum_{i=1}^8 \vert  \alpha_i  \vert^2  =1$.
All such states -- in particular, the ``original''
GHZ state
$\vert \Upsilon_1 \rangle$
or any coherent superposition such as
$
\vert \Omega \rangle =
 \frac{1}{2\sqrt{2}} \sum_{i=1}^8 \vert \Upsilon_i \rangle =
\frac{1}{2} \begin{pmatrix} 1,1,1,1,0,0,0,0 \end{pmatrix}^\intercal
$
--
produce a complete discord or contradiction with any classical prediction~(\ref{2020-ghz-ghlmcont})~\cite[Appendix]{mermin1}.
The particular form of those discords will be discussed later.

\begin{figure}[h]
\begin{center}
\includegraphics{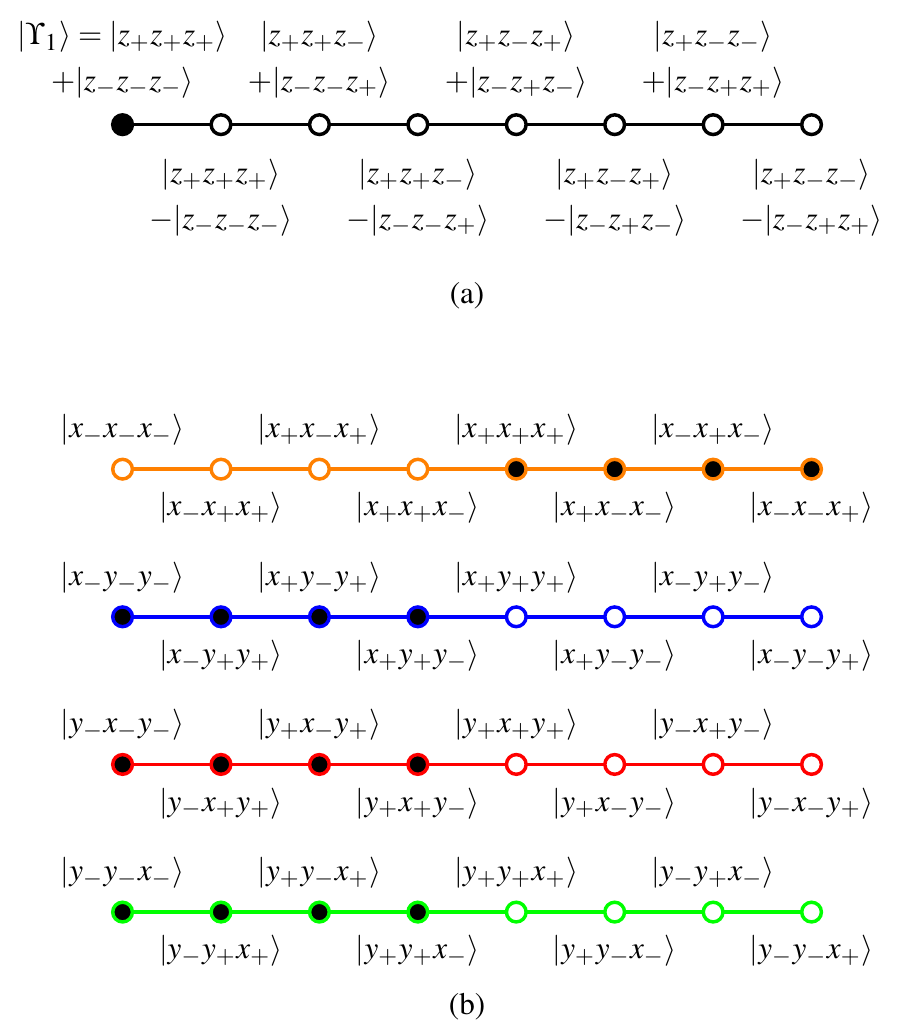}
\end{center}
\caption{\label{2020-f-ghz-contextqm}
Hypergraphs representing  (a) GHZ states forming a single quantum mechanical GHZ state context
represented by the GHZ basis of the eigenstates of $\sigma_z \sigma_z \sigma_z$,
and (b) the four nonintertwining GHZ contexts in the four bases of the eigenstates of
$\sigma_y \sigma_y \sigma_x$, $\sigma_y \sigma_x \sigma_y$, $\sigma_x \sigma_y \sigma_y$,  and $\sigma_x \sigma_x \sigma_x$
with eight atoms each.
Filled circles indicate states ``allowed'' by---that is, are equal to or
occur in the coherent superposition of---the original GHZ game state
$\vert \Upsilon_1 \rangle = \vert z_+z_+z_+ \rangle  + \vert z_-z_-z_- \rangle$
as enumerated in Eqs.~(\ref{2020-ghz-evsisjsk}) and~(\ref{2020-ex-rew-ghz}),
yielding a discord with classical means and predictions.
A comparison of the classical $\{x,x,x\}$ context drawn on top of Figure~\ref{2020-f-ghz-contextconf}(b)
yields an ``inverted'' situation:
the ``allowed'' states are classically ``disallowed'', and vice versa.
}
\end{figure}

What has happened to the four classical contexts ``in transit'' to the quantum configurations?
Not very much: the chosen state or share is in a single quantum context
depicted in Figure~\ref{2020-f-ghz-contextqm}(a).
This context can be ``expanded'' into the four known classical contexts discussed earlier by rewriting its atoms
in terms of the eigenstates (eigenvalues) relevant observables (operators)
$\sigma_y \sigma_y \sigma_x$,
$\sigma_y \sigma_x \sigma_y$,
$\sigma_x \sigma_y \sigma_y$, and
$\sigma_x \sigma_x \sigma_x$.
The quantum GHZ argument never ``leaves'' those contexts.

Indeed, one might maintain that the single context of GHZ states associated with the GHZ basis, whose atomic propositions
have a faithful orthogonal representation~\cite{lovasz-79,lovasz-89,Grtschel1993,Portillo-2015}
enumerated in Eq.~(\ref{2020-ghz-evsisjsk}), is ``viewed'' from different ``projective angles''
represented in terms of the eigenstates of $\sigma_y \sigma_y \sigma_x$, $\sigma_y \sigma_x \sigma_y$, $\sigma_x \sigma_y \sigma_y$,  and $\sigma_x \sigma_x \sigma_x$
as depicted in Figure~\ref{2020-f-ghz-contextqm}(b)
by the four contexts enumerated in Eq.~(\ref{2020-ghz-sigmas-contextop}).
This amounts to a state partitioning or filters~\cite{DonSvo01,svozil-2002-statepart-prl,svozil-2005-ko}.
Explicitly, the eigenvectors are
\begin{equation}
\begin{aligned}
& \vert x_+  x_+  x_+ \rangle = \begin{pmatrix}1, 1, 1, 1, 1, 1, 1, 1\end{pmatrix}^\intercal , \\
& \vert x_+  x_-  x_- \rangle = \begin{pmatrix}1, -1, -1, 1, 1, -1, -1, 1\end{pmatrix}^\intercal , \\
& \vert x_-  x_+  x_- \rangle = \begin{pmatrix}1, -1,  1, -1, -1, 1, -1, 1\end{pmatrix}^\intercal , \\
& \vert x_-  x_-  x_+ \rangle = \begin{pmatrix}1, 1, -1, -1, -1, -1, 1, 1\end{pmatrix}^\intercal , \\
& \vert x_-  y_-  y_- \rangle = \begin{pmatrix}1, -i, -i, -1, -1, i, i, 1\end{pmatrix}^\intercal , \\
& \vert x_-  y_+  y_+ \rangle = \begin{pmatrix}1, i, i, -1, -1, -i, -i, 1\end{pmatrix}^\intercal , \\
& \vert x_+  y_-  y_+ \rangle = \begin{pmatrix}1, i, -i, 1, 1, i, -i, 1\end{pmatrix}^\intercal , \\
& \vert x_+  y_+  y_- \rangle = \begin{pmatrix}1, -i, i, 1, 1, -i, i, 1 \end{pmatrix}^\intercal , \\
& \vert y_-  x_-  y_- \rangle = \begin{pmatrix}1, -i, -1, i, -i, -1, i, 1\end{pmatrix}^\intercal , \\
& \vert y_-  x_+  y_+ \rangle = \begin{pmatrix}1, i, 1, i, -i, 1, -i, 1\end{pmatrix}^\intercal , \\
& \vert y_+  x_-  y_+ \rangle = \begin{pmatrix}1, i, -1, -i, i, -1, -i, 1\end{pmatrix}^\intercal , \\
& \vert y_+  x_+  y_- \rangle = \begin{pmatrix}1, -i, 1, -i, i, 1, i, 1 \end{pmatrix}^\intercal , \\
& \vert y_-  y_-  x_- \rangle = \begin{pmatrix}1, -1, -i, i, -i, i, -1, 1\end{pmatrix}^\intercal , \\
& \vert y_-  y_+  x_+ \rangle = \begin{pmatrix}1, 1, i, i, -i, -i, 1, 1\end{pmatrix}^\intercal , \\
& \vert y_+  y_-  x_+ \rangle = \begin{pmatrix}1, 1, -i, -i, i, i, 1, 1\end{pmatrix}^\intercal , \\
& \vert y_+  y_+  x_- \rangle = \begin{pmatrix}1, -1, i, -i, i, -i, -1, 1\end{pmatrix}^\intercal
.
\end{aligned}
\end{equation}

Another, algebraic, way of perceiving this is in terms of the mutual commutativity of
the four operators
$\sigma_y \sigma_y \sigma_x$, $\sigma_y \sigma_x \sigma_y$, $\sigma_x \sigma_y \sigma_y$,  and $\sigma_x \sigma_x \sigma_x$
in Eq.~(\ref{2020-ghz-sigmas-contextop}), which thus can be written as real-valued functions of
a maximal operator~\cite[\S~84, p.~171,172]{halmos-vs}
\begin{equation}
\textsf{\textbf{R}} = \sum_{i=1}^8 \lambda_i \vert \Upsilon_i \rangle \langle  \Upsilon_i \vert ,
\label{2020-ghz-mo}
\end{equation}
with mutually different $\lambda_i$ and $\vert \Upsilon_i \rangle$ defined in Eq.~(\ref{2020-ghz-evsisjsk}).
As can be readily read off from Table~\ref{2020-hardy-t-eveeva}, and with
$f_{yyx}(\textsf{\textbf{R}})=\sigma_y \sigma_y \sigma_x$,
$f_{yxy}(\textsf{\textbf{R}})=\sigma_y \sigma_x \sigma_y$,
$f_{xyy}(\textsf{\textbf{R}})=\sigma_x \sigma_y \sigma_y$,  and
$f_{xxx}(\textsf{\textbf{R}})=\sigma_x \sigma_x \sigma_x$
these four functions need to obey
$
-f_{yyx}(\lambda_1)=
f_{yyx}(\lambda_2)=
-f_{yyx}(\lambda_3)=
f_{yyx}(\lambda_4)=
f_{yyx}(\lambda_5)=
-f_{yyx}(\lambda_6)=
f_{yyx}(\lambda_7)=
-f_{yyx}(\lambda_8)=
-f_{yxy}(\lambda_1)=
f_{yxy}(\lambda_2)=
f_{yxy}(\lambda_3)=
-f_{yxy}(\lambda_4)=
-f_{yxy}(\lambda_5)=
f_{yxy}(\lambda_6)=
f_{yxy}(\lambda_7)=
-f_{yxy}(\lambda_8)=
-f_{xyy}(\lambda_1)=
f_{xyy}(\lambda_2)=
f_{xyy}(\lambda_3)=
-f_{xyy}(\lambda_4)=
f_{xyy}(\lambda_5)=
-f_{xyy}(\lambda_6)=
-f_{xyy}(\lambda_7)=
f_{xyy}(\lambda_8)=
f_{xxx}(\lambda_1)=
-f_{xxx}(\lambda_2)=
f_{xxx}(\lambda_3)=
-f_{xxx}(\lambda_4)=
f_{xxx}(\lambda_5)=
-f_{xxx}(\lambda_6)=
f_{xxx}(\lambda_7)=
-f_{xxx}(\lambda_8)=
1$.

\subsection{Verification of the original quantum GHZ game}

Let us review our solution to the original GHZ game.
It can be won with
quantum resources from the GHZ basis
enumerated in (\ref{2020-ghz-context}) and (\ref{2020-ghz-evsisjsk}).
If the prisoners share a particular quantum state representable by Eq.~(\ref{2020-ghz-psiGHZ}), namely
$\vert \Upsilon_1 \rangle = \frac{1}{\sqrt{2}}
\left(
\vert {z_+}{z_+}{z_+} \rangle +
\vert {z_-}{z_-}{z_-} \rangle
\right)$
prepared in one direction, say $z$,
prior to being separated, this goal can be achieved.
Because the prisoners measure, on their respective sides and constituents particles
(the share includes three constituents because it is a three-partite state),
 $\vert \Upsilon_1 \rangle$ either in the $x$ basis
$\vert x_\pm \rangle  = \frac{1}{\sqrt{2}}
\left(
\vert {z_+} \rangle \pm
\vert {z_-} \rangle
\right)  = \frac{1}{\sqrt{2}}
\begin{pmatrix}
1,\pm 1
\end{pmatrix}
$
[conversely,
$\vert {z_\pm} \rangle  = \frac{1}{\sqrt{2}}
\left(
\vert x_+ \rangle \pm
\vert x_- \rangle
\right)
${\large]}
whenever their slip says ``$x$'',
or   in the $y$ basis
$\vert y_\pm \rangle  = \frac{1}{\sqrt{2}}
\left(
\vert {z_+} \rangle \pm
i
\vert {z_-} \rangle
\right) = \frac{1}{\sqrt{2}}
\begin{pmatrix}
1,\pm i
\end{pmatrix}
$
[conversely,
$\vert {z_+} \rangle  = \frac{1}{\sqrt{2}}
\left(
\vert y_+ \rangle +
\vert y_- \rangle
\right)$
and $\vert {z_-} \rangle  = -\frac{i}{\sqrt{2}}
\left(
\vert y_+ \rangle +
\vert y_- \rangle
\right)
${\large]}
whenever their slip says ``$y$''.

That this goal can be perfectly---ideally at all instances and configurations---reached
can be demonstrated by rewriting the
GHZ state $\vert \Upsilon_1 \rangle $
in terms of the basis vectors of the four configurations or contexts,
one of which is singled out (but not disclosed to the prisoners) by the prison ward
[see also the filled states in the contexts depicted in Figure~\ref{2020-f-ghz-contextqm}(b)]:
\begin{equation}
\begin{aligned}
\vert \Upsilon_1 \rangle
&=  \frac{1}{2}
\left(
\vert x_+  x_+  x_+  \rangle +
\vert x_+  x_-  x_-  \rangle +
\vert x_-  x_+  x_-  \rangle +
\vert x_-  x_-  x_+  \rangle
\right)  \\
&=  \frac{1}{2}
\left(
\vert x_-  y_-  y_-  \rangle +
\vert x_-  y_+  y_+  \rangle +
\vert x_+  y_-  y_+  \rangle +
\vert x_+  y_+  y_-  \rangle
\right)  \\
&=  \frac{1}{2}
\left(
\vert y_-  x_-  y_-  \rangle +
\vert y_-  x_+  y_+  \rangle +
\vert y_+  x_-  y_+  \rangle +
\vert y_+  x_+  y_-  \rangle
\right)  \\
&=  \frac{1}{2}
\left(
\vert  y_-  y_-  x_-  \rangle +
\vert  y_-  y_+  x_+  \rangle +
\vert  y_+  y_-  x_+  \rangle +
\vert  y_+  y_+  x_-  \rangle
\right)
.
\end{aligned}
\label{2020-ex-rew-ghz}
\end{equation}
For every individual experimental run,
the share ``produces'' or ``selects'' one of the four outcomes from the coherent superpositions.

Note that there is a ``hidden cost''  in this game:
to make sure that the quantum $xxx$ realization conforms with the other three choices  $xyy$, $yxy$, and $yyx$
of the ward, one needs to either believe in this counterfactual supposition
or make additional experiments to verify the latter three.
But more experimental runs are added which are ``hidden'' away by acknowledging
that first checks of $xyy$, $yxy$, and $yyx$ have to be performed; and, as stated in Ref.~\cite[p.~517]{panbdwz},
{\it ``if the results obtained are in agreement with the predictions for a GHZ
state, then for an $xxx$ experiment, our expectations
using a local-realist theory are exactly the opposite of our expectations
using quantum physics.''}
Confidence about whether one is dealing with a GHZ state $\vert \Upsilon_1 \rangle $
(by checking the outcomes of the $xyy$, $yxy$, and $yyx$ and finally $xxx$ configurations)
requires additional experimental runs.

The GHZ scenario can be interpreted as some nonlocal form of Hardy-type
true-implies-true scenario with ``maximal aperture'' $\frac{\pi}{2}$.
This can, for example, be achieved by
serial composition~\cite{svozil-2018-whycontexts} of such true-implies-true gadgets~\cite{2018-minimalYIYS};
a procedure already performed by Kochen and Specker~\cite{kochen1,specker-ges}, as they move from $\Gamma_1$ to $\Gamma_2$.

\section{Generalization of GHZ games}

Table~\ref{2020-hardy-t-eveeva} enumerates the eight variations of GHZ games
with viable winning strategies for the prisoners.
Those strategies are based on modified GHZ states as assets.
By the earlier parity argument, it is again impossible for the prisoners employing classical means
to win for all such paper slip configurations the ward might issue.
These variations can be translated into experimental protocols analogous to the ones executed in Refs.~\cite{PhysRevLett.82.1345,panbdwz}.

Suppose, for the sake of an example, the prison ward requires the respective products of
$xyy$, $yxy$, and $yyx$  to turn out positive and $xxx$ to turn out negative.
This can be won by the prisoners with a $\vert \Upsilon_2 \rangle = \frac{1}{\sqrt{2}}\left( \vert z_+z_+z_+ \rangle  - \vert z_-z_-z_- \rangle \right) $ share.

Or suppose, for the sake of another example, the prison ward requires the respective products of
$xyy$  to turn out positive and $yyx$, $yxy$, and $xxx$ to turn out negative.
The prisoners can win this with a $\vert \Upsilon_8 \rangle = \frac{1}{\sqrt{2}}\left( \vert z_+z_-z_- \rangle  - \vert z_-z_+z_+ \rangle \right) $ share.
The respective game components are
\begin{equation}
\begin{aligned}
\vert \Upsilon_8 \rangle
&=  -\frac{1}{2}
\left(
\vert x_+x_+x_- \rangle +
\vert x_+x_-x_+ \rangle -
\vert x_-x_-x_- \rangle -
\vert x_-x_+x_+ \rangle
\right)  \\
&=  \frac{1}{2}
\left(
\vert x_-y_+y_- \rangle +
\vert x_-y_-y_+ \rangle -
\vert x_+y_-y_- \rangle -
\vert x_+y_+y_+ \rangle
\right)  \\
&=  -\frac{i}{2}
\left(
\vert y_-x_-y_- \rangle -
\vert y_+x_+y_- \rangle +
\vert y_-x_+y_+ \rangle -
\vert y_+x_-y_+ \rangle
\right)  \\
&=  -\frac{i}{2}
\left(
\vert y_-y_-x_- \rangle -
\vert y_+y_-x_+ \rangle +
\vert y_-y_+x_+ \rangle -
\vert y_+y_+x_- \rangle
\right)
.
\end{aligned}
\label{2020-ex-rew-ghz8}
\end{equation}

Because of symmetry, we may expect that not much changes regarding gambling if we permute $\sigma_1 \leftrightarrow \sigma_2$. Here we need to form a
modified GHZ basis serving as eigenstates, with the first ``$1$'' component of the vectors of the original GHZ basis substituted by the imaginary unit ``$i$''.
With those eigenstates Table~\ref{2020-hardy-t-eveeva} is reproduced.

\section{Games that can be won by classical but not by quantum GHZ means}
\label{2020-ghz-gtcbwbcbnbqm}

It should also be mentioned what the prisoners cannot achieve with GHZ type quantum resources.
Table~\ref{2020-hardy-t-eveeva} does not allow a
perfect winning strategy for the aforementioned games in which the ward requests from the prisoners to produce only positive or negative outcomes.
They can win these games with the classical strategies or shares mentioned earlier.
Table~\ref{2020-hardy-t-eveevac} enumerates the configurations
and some of the (eight per configuration) classical winning strategies.

\begin{table}[t]
 \begin{center}
 \caption{Enumeration of the eight configurations that cannot be won with quantum GHZ type shares, and one (of the eight per configuration) classical winning strategies.
\label{2020-hardy-t-eveevac}}
 \begin{tabular}{ c c c c c c c c c c c c }
\toprule
$1^\text{st}$ prisoner&$2^\text{nd}$ prisoner&$3^\text{rd}$ prisoner&    $yyx$ & $yxy$ &  $xyy$  & $xxx$
\\
\midrule
$x=-1$ & $x=-1$ & $x=-1$&  {\Large -} & {\Large -} &  {\Large -} & {\Large -}\\
$y=-1$ & $y=-1$ & $y= -1$\\
$x=-1$ & $x=-1$ & $x=+1$&   {\Large -} & {\Large -} &  {\Large +} & {\Large +}\\
$y=-1$ & $y=+1$ & $y= -1$\\
$x=-1$ & $x=-1$ & $x=-1$&   {\Large -} & {\Large +} &  {\Large +} & {\Large -}\\
$y=-1$ & $y=-1$ & $y= +1$\\
$x=-1$ & $x=-1$ & $x=+1$&   {\Large -} & {\Large +} &  {\Large -} & {\Large +}\\
$y=-1$ & $y=+1$ & $y= +1$\\
$x=-1$ & $x=-1$ & $x=-1$&   {\Large +} & {\Large -} &  {\Large +} & {\Large -}\\
$y=-1$ & $y=+1$ & $y= -1$\\
$x=-1$ & $x=-1$ & $x=+1$&   {\Large +} & {\Large -} &  {\Large -} & {\Large +}\\
$y=-1$ & $y=-1$ & $y= -1$\\
$x=-1$ & $x=-1$ & $x=-1$&   {\Large +} & {\Large +} &  {\Large -} & {\Large -}\\
$y=-1$ & $y=+1$ & $y= +1$\\
$x=+1$&$x=+1$&$x=+1$& {\Large +} & {\Large +} &  {\Large +} & {\Large +}\\
$y=+ 1$ & $y=+1$ & $y= + 1$   \\
\botrule
\end{tabular}
 \end{center}
 \end{table}

\section{Classical contextuality or context dependence}

An examination of the subgraph of the GHZ logic hypergraph, depicted in Figure~\ref{2020-f-ghz-contextqm}(b) ``covered'' by $\vert \Upsilon_1 \rangle$
shows that this (sub)logic has a separating set of $8^4$ admissible two-valued states
satisfying completeness (every context contains a red or 1 value) and exclusivity (every context contains only one red or 1 value, all other vertices are green or 0 value).

\begin{figure}[h]
\begin{center}
\includegraphics[width=0.5\textwidth]{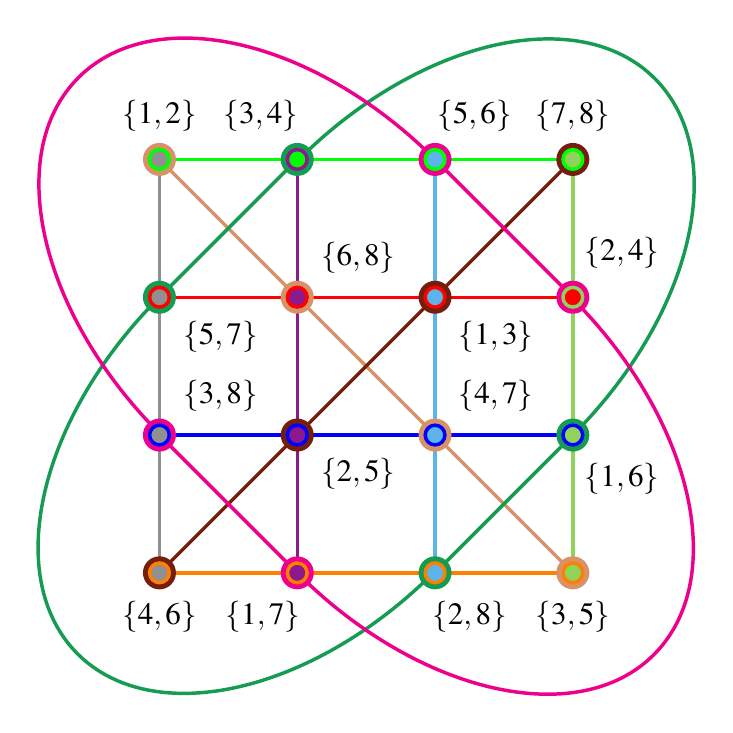}
\end{center}
\caption{\label{2020-f-ghz-contextpl}
Hypergraphs representing 12 tightly intertwined ``tightened'' GHZ contexts in a partition logic representation.
}
\end{figure}

\begin{figure}[h]
\begin{center}
\begin{tabular}{ c c c c}
\includegraphics[width=0.2\textwidth]{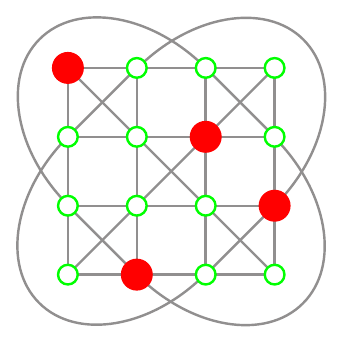}
&
\includegraphics[width=0.2\textwidth]{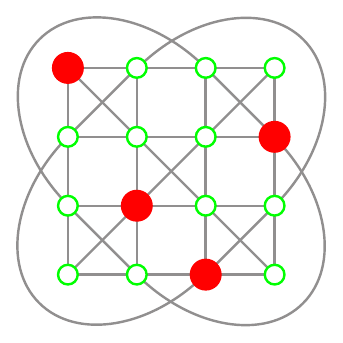}
&
\includegraphics[width=0.2\textwidth]{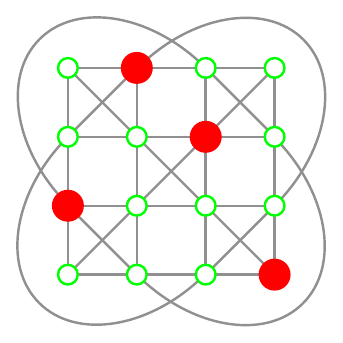}
&
\includegraphics[width=0.2\textwidth]{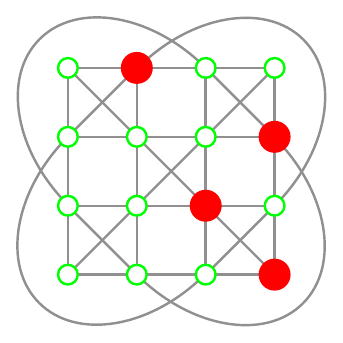}
\\
\includegraphics[width=0.2\textwidth]{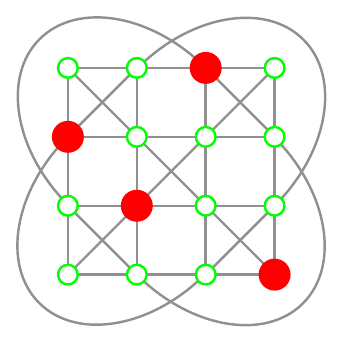}
&
\includegraphics[width=0.2\textwidth]{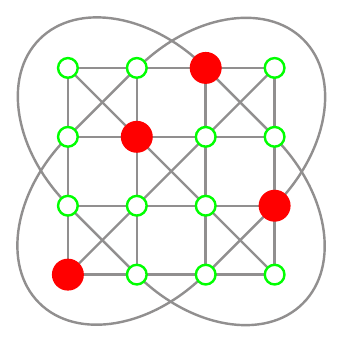}
&
\includegraphics[width=0.2\textwidth]{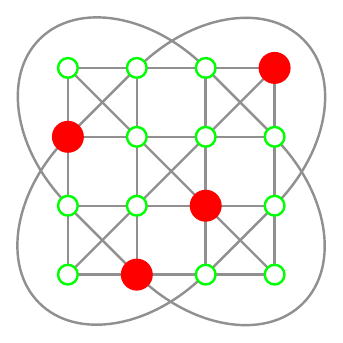}
&
\includegraphics[width=0.2\textwidth]{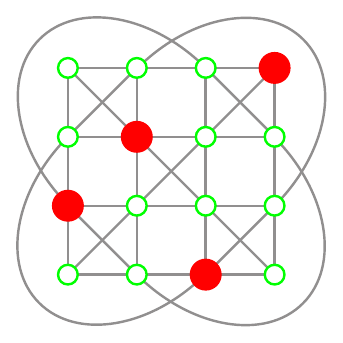}
\end{tabular}
\end{center}
\caption{\label{2020-f-ghz-contextghz2vs}
Enumeration of the eight two-valued states of the ``tightened'' GHZ logic.
}
\end{figure}

For the sake of demonstration, this propositional structure can further be
``tightened'' by introducing four additional ``vertical'' and four additional ``diagonal'' contexts as depicted in Figure~\ref{2020-f-ghz-contextpl}.
The resulting tightened GHZ (sub)logic comprising 12 contexts supports eight two-valued states which are separating any two different propositions.
They are enumerated in Figure~\ref{2020-f-ghz-contextghz2vs}.
(In this presentation the two-valued states appear like $0-1$ entries in permutation matrices.)
Consequently, this configuration has a classical
representation by a partition logic $\{1,2,\ldots ,8\}$ of eight elements, as drawn in Figure~\ref{2020-f-ghz-contextpl}.

From this partition logic, given by
\begin{equation}
\begin{aligned}
&\{ \{
\{1,2\},\{3,4\},\{5,6\},\{7,8\}
\},\{
\{5,7\},\{6,8\},\{1,3\},\{2,4\}
\},\\
&\{
\{3,8\},\{2,5\},\{4,7\},\{1,6\}
\},\{
\{4,6\},\{1,7\},\{2,8\},\{3,5\}
\},\\
&\{
\{1,2\},\{6,8\},\{4,7\},\{3,5\}
\},\{
\{7,8\},\{1,3\},\{2,5\},\{4,6\}
\},\\
&\{
\{5,6\},\{2,4\},\{3,8\},\{1,7\}
\},\{
\{3,4\},\{5,7\},\{1,6\},\{2,8\}
\}
\}
,
\end{aligned}
\label{2020-ghz-e-pl}
\end{equation}

one can immediately derive the classical probability distributions,
as well as a classical winning strategy for a GHZ-type game that is context-dependent but requires knowledge of the configuration or context involved.
All that is needed is the identification of the respective vertices in the original configuration and the partition logic
(any other variation within the horizontal contexts would also suffice);
e.g.,
$\{1,2\} \mapsto x_+x_+x_+$,
$\{3,4\} \mapsto x_+x_-x_-$, $\ldots$,
$\{2,8\} \mapsto y_+y_+x_-$.
The associated generalized urn model might match important aspects of the quantum mechanical predictions such as expectation values and distributions of order one, two, and three.

A typical experimental run would comprise a generalized urn ``loaded'' with (triples of identical) balls of eight types and four colors, associated with
the ward's potential choice of configurations or contexts.
Upon drawing such a ball triple the prisoners take their share of the triple---exactly one ball from the triple per prisoner---and, in accordance with the ward's choice,
read the pairs of numbers (from one to eight) in that respective color. Their choice must then be their position of the identified answer.
More explicitly, if they read ``$3,4$'' in the color associated with the first context,
the first prisoner writes or shouts ``$+$''1,
the second prisoner writes or shouts ``$-$''1,
the third prisoner writes or shouts ``$-$''1.
As has been mentioned earlier, any such value assignment is contextual in at least one observable $x$ or $y$; that is, for example, the first prisoner writes or shouts
different values for $x$ or $y$ for different contexts.

\section{Stranger-than-quantum GHZ games}

\begin{figure}[htb]
\begin{center}
\begin{tabular}{ c }
\includegraphics{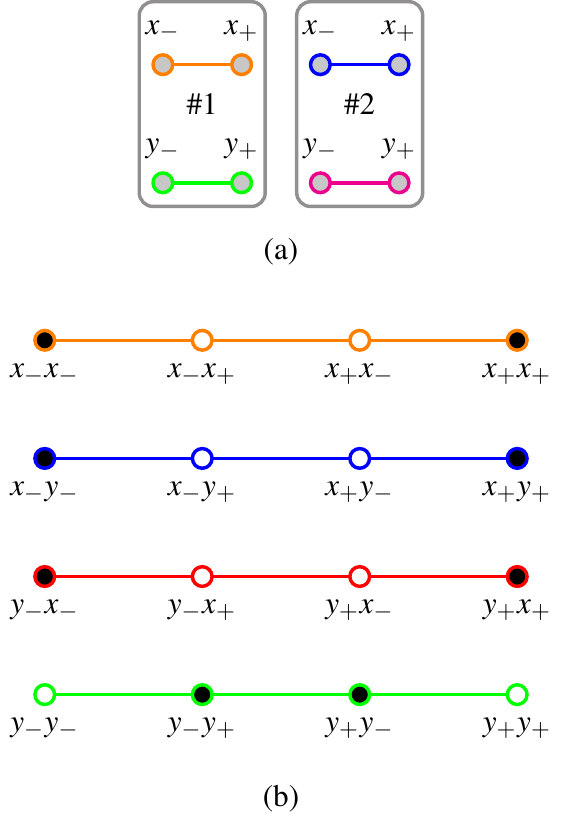}
\end{tabular}
\end{center}
\caption{\label{2020-f-ghz-contextconfstranger}
Hypergraphs representing
(a) the four disconnected classical single-particle contexts---two contexts per particle---representing the observables of the first and second particle, respectively;
(b) the four separated, nonintertwined two-partite contexts, with four atoms each, that are the Cartesian products of these single-particle contexts.
Filled circles indicate states which are involved in a stranger-than-GHZ game
requesting positivity for all products  $xx=xy=yx=+1$ and  negativityfor $yy=-1$.
}
\end{figure}

What if the prisoners would be given a ``magic filter'' or share capacity---say, a ``nonlocal'' (NL) or Popescu Rohrlich (PR) box~\cite{pop-rohr,svozil-krenn}---to go beyond quantum capacities of the GHZ game?
Suppose, for instance, that there are merely two prisoners involved, and four two-partite configurations or contexts
$xx$,
$xy$,
$yx$, and
$yy$; with the ward requesting as goal to achieve a negative product for $y\cdot y$
and else---that is, for $x\cdot x$, $x\cdot y$, and $y\cdot x$---positive products, as depicted in Figure~\ref{2020-f-ghz-contextconfstranger}.
Then again, by a parity argument such configurations cannot exist classically, because the observables $x$ and $y$ on either side contribute
to the overall product in pairs $x^2$ and $y^2$ which are always positive for real-valued $x$ and $y$;
but the ward requests a negative joint product
$\underbrace{(x\cdot x)}_{+1}\underbrace{(x\cdot y)}_{+1}\underbrace{(y\cdot x)}_{+1}\underbrace{(y\cdot y)}_{-1}=\left[ x^2 y^2\right]^2=(xy)^4=-1$
for real valued $x$ and $y$.

For a proof that this configuration has no quantum realization
we take (overline indicates complex conjugation)
$\vert x_+ \rangle =\begin{pmatrix} x_{+,1} , x_{+,2} \end{pmatrix}^\intercal$,
$\vert x_- \rangle =\begin{pmatrix} \overline{x}_{+,2} , -\overline{x}_{+,1}  \end{pmatrix}^\intercal$,
$\vert y_+ \rangle =\begin{pmatrix} y_{+,1} , y_{+,2} \end{pmatrix}^\intercal$,
$\vert y_- \rangle =\begin{pmatrix} \overline{y}_{+,2} , -\overline{y}_{+,1}  \end{pmatrix}^\intercal$,
and $\vert \psi \rangle = \begin{pmatrix} \psi_1,\psi_2,\psi_3,\psi_4 \end{pmatrix}^\intercal$.
The stranger-than-GHZ game rules require that
$
\langle  \psi  \vert  x_+ x_+ \rangle =
\langle  \psi  \vert  x_- x_- \rangle =
\langle  \psi  \vert  x_+ y_+ \rangle =
\langle  \psi  \vert  x_- y_- \rangle =
\langle  \psi  \vert  y_+ x_+ \rangle =
\langle  \psi  \vert  y_- x_- \rangle =
\langle  \psi  \vert  y_+ y_- \rangle =
\langle  \psi  \vert  y_- y_+ \rangle = 0$.
Without loss of generality we may fix $x_{+,2}=0$, and, because of symmetry, $y_{+,1} = y_{+,2} = 1/\sqrt{2}$,
which admits only the trivial solution  $\vert \psi \rangle = \begin{pmatrix}0,0,0,0 \end{pmatrix}^\intercal$.

The situation is ``contextual'' as any realization involves context-dependence of either the $x$ or the $y$-observables.
Suppose the prisoners have at their disposal a NL PR box of the following capacity:
if $i_1$ and $i_2$ stand for the inputs of the box, and $o_1$ and $o_2$ for its outputs, then
$
i_1 i_2 = o_1 \oplus o_2 = o_1 + o_2 \text{ mod }2$.
Now if both prisoners ``translate their in- and outputs'' as follows:
they input 0 on an "x" input, and 1 on a "y" input,
as well as identify  ``$+$''1 with output 0 and ``$-$''1 with output 1, respectively.
Then, if the in- and outputs are ordered by prisoner,
\begin{equation}
\begin{aligned}
x,x & \mapsto   0 \cdot 0 = 0 \stackrel{\text{NL PR}}{\xrightarrow{\hspace*{6mm}}} \{0 \oplus 0 \text{  or  } 1 \oplus 1\} \mapsto   \{++,--\}     ,\\
x,y & \mapsto   0 \cdot 1 = 0 \stackrel{\text{NL PR}}{\xrightarrow{\hspace*{6mm}}} \{0 \oplus 0 \text{  or  } 1 \oplus 1\} \mapsto   \{++,--\}     ,\\
y,x & \mapsto   1 \cdot 0 = 0 \stackrel{\text{NL PR}}{\xrightarrow{\hspace*{6mm}}} \{0 \oplus 0 \text{  or  } 1 \oplus 1\} \mapsto   \{++,--\}     ,\\
y,y & \mapsto   1 \cdot 1 = 1 \stackrel{\text{NL PR}}{\xrightarrow{\hspace*{6mm}}} \{0 \oplus 1 \text{  or  } 1 \oplus 0\} \mapsto   \{+-,-+\}     .
\end{aligned}
\end{equation}
This perfect winning strategy can be further specified~\cite{PhysRevA.71.022101} by
the joint outcome probabilities
$P \left( o_1,o_2 \vert i_1,i_2 \right) = \frac{1}{2}$ for $i_1 i_2 = o_1 \oplus o_2 = o_1 + o_2$ and vanishing otherwise.
For a game requiring
$\underbrace{(x\cdot x)}_{-1}\underbrace{(x\cdot y)}_{-1}\underbrace{(y\cdot x)}_{-1}\underbrace{(y\cdot y)}_{+1}=-1$
one of the prisoners would need to switch the output behaviour $+ \longleftrightarrow -$.
Indeed the NL PR box appears to be a direct realization of a winning strategy  for the stranger-than-GHZ game.

\section{Some afterthoughts}

From a quantum logical point of view the GHZ argument differs
from the KS argument in its fairly simplistic geometric and algebraic structure involving only four isolated contexts
instead of a collection of intertwining contexts.
Indeed, it might be amazing how much quantum-versus classical discord can be gained from such configuration.
Unlike the KS configuration which is based on the total absence of two-valued states interpretable as classical truth assignments,
the observables involved in GHZ allow a plethora (namely $8^4$) of such states which are separating~\cite[Theorem~0]{kochen1,specker-ges}.

The strength of the GHZ argument lies not in the scarcity of two-valued states
but in the proper---that is, suitable for, and adapted to, the particular task---reduction or ``filtering'' the respective eigenstates, which serve as elements of the contexts,
by the state preparation or selection.
Thereby some nonclassical features can be claimed relative to the assumptions.

This is not dissimilar to ``Hardy-type'' configurations in which a particular particle or a multi-partite state is prepared and---by the classical interpretation
of a configuration of observables forming intertwining contexts---``viewed'' by another observable or proposition~\cite{svozil-2020-hardy}.
In particular, if such a configuration with a true-implies-true set of two-valued states (TITS)~\cite{2018-minimalYIYS}
has ``orthogonal arperture'' (that is, orthogonality between the prepared and the measured state)
the classical prediction is in direct contradiction with the quantum expectation of nonoccurence because of orthogonality.
This has already been realized by KS as subgraphs of their $\Gamma_2$---they render the orthogonal aperture by a serial composition of TITS (their $\Gamma_1$)
with less than $\frac{\pi}{2}$ between the states prepared and measured~\cite{kochen1,specker-ges}.

Another difference to KS and Hardy-type arguments comes from the fact that its predictions are obtained by multiplying nonvanishing values in $\{-1,+1\}$
of the experimental outcomes instead of working with dispersionless $\{0,1\}$ two-valued states which include the value $0$.
Taking nonvanishing values $\{-1,+1\}$ preserves much more information and structure about the joint observables than in the $\{0,1\}$ case.
Because while for a single two-state particle
any $\{-1,+1\}$-observable $\textsf{\textbf{A}}$ can be considered an affine transformation
of the two-valued state $s \in \{0,1\}$ by $\textsf{\textbf{A}} = 2 s - 1$
(that corresponds to the negative of a quantum Householder transformation
$\textsf{\textbf{U}}
=
\mathbb{1}- 2  \vert {\bf x} \rangle \langle  {\bf x}  \vert$ for a unit vector $\vert {\bf x} \rangle$),
there is no loss of information due to this linear transformation,
this is in general not true in the multi-partite situation; in particular, with three particles involved:
whereas in the  $\{0,1\}$ case all products vanish except in the case for which the outcome is 1 on all three particles simultaneously,
the products of possible  $\{-1,+1\}$ cases show a much more balanced partitioning:
parity dictates that there is equidistribution (50:50) of joint negative and positive outcomes.
Formally, in terms of entropy, the  $\{0,1\}$  case results in
$H_{\{0,1\}^3} = -\frac{1}{8} \log_2 \frac{1}{8} -  \frac{7}{8} \log_2 \frac{7}{8}   \approx 0.54$
as compared to the  $\{-1,+1\}$ case which yields
$H_{\{-1,+1\}^3} = -\frac{1}{2} \log_2 \frac{1}{2} -  \frac{1}{2} \log_2 \frac{1}{2}  =1$.
Therefore, the latter three-partite $\{-1,+1\}$ outcome case results in an average rate of ``information'' inherent in the variable's potential outcomes
which is almost twice as high as in the  $\{0,1\}$  case.
One straightforward way of enhancing the ``contrast'' or violation between classical and quantum predictions would be to
consider Operators with eigenvalues $\{-\lambda,+\lambda\}$ larger that in the ones chosen for the GHZ argument; that is, $1\ll \lambda $.
The respective observables correspond to operators whose spectral sum includes the same orthogonal projection operators as the operators~(\ref{2020-ghz-sigmas-contextop})
involved in the GHZ case. These orthogonal projection operators are
formed by the dyadic products of the vectors~(\ref{2020-ghz-evsisjsk}).

This review is the continuation of an ongoing effort~\cite{svozil-2017-b,svozil-2020-hardy}
to transcribe and (de)construct well-known arguments from the foundations of quantum mechanics into a logico-algebraic context.
The operators of the GHZ argument had to be delineated in terms of their spectral form, revealing the propositional structure in terms of the
orthogonal projection operators. Since the argument is state dependent, the valuations of these elementary propositional observables
had to be taken for the particular state involved.
My aim has been to obtain insights into the structural properties of the arguments,
as well as transparent and effective presentations and derivations
of the predictions and probabilities supported by the respective configurations.

Following the aforementioned effort, the conceptual take-home message of the paper is this: the GHZ argument can be readily ``embedded'' into,
or translated and presented, in terms of the quantum logical framework of Birkhoff and von Neumann~\cite{birkhoff-36}
as well as partial algebras~\cite{kochen2,kochen3,specker-ges} and orthomodular structures~\cite{kalmbach-83,pulmannova-91}.
From a formal point of view, this allows its uniform presentation and comparison to ``competing'' and related arguments,
such as Hardy-type or KS arguments.
From an empirical point of view, the operator-based approach facilitates operational extensions and generalizations of the original argument.

\backmatter

\bmhead{Acknowledgments}

I kindly acknowledge discussions with and corrections by Peter Morgan, as well as a hint by Anne Broadbent with regards to realizations of stranger-than-quantum GHZ games by NL PR boxes.
All misconceptions and errors are mine.

This research was funded in whole, or in part, by the Austrian Science Fund (FWF), Project No. I 4579-N. For the purpose of open access, the author has applied a CC BY public copyright licence to any Author Accepted Manuscript version arising from this submission.






\end{document}